\documentclass{article}

\usepackage{epsfig}
\usepackage{subfigure}
\usepackage{wrapfig}

\usepackage{amsmath}
\usepackage{amssymb}
\usepackage{cite}
\usepackage{array}

\usepackage{fancyhdr}
\newcommand{\msc}{\Phi}

\numberwithin{equation}{section}
\numberwithin{figure}{section}

\newcommand{\beqs}{\begin{equation*}}
\newcommand{\beq}{\begin{equation}}

\newcommand{\eeqs}{\end{equation*}}
\newcommand{\eeq}{\end{equation}}

\newcommand{\beqas}{\begin{eqnarray*}}
\newcommand{\beqa}{\begin{eqnarray}}

\newcommand{\eeqas}{\end{eqnarray*}}
\newcommand{\eeqa}{\end{eqnarray}}

\newcommand{\eq}[2]{\begin{equation} #1 \label{#2} \end{equation}}

\newcommand{\meq}[2]{\begin{multline} #1 \label{#2} \end{multline}}

\newcommand{\eps}{\varepsilon}
\newcommand{\al}{\alpha}
\newcommand{\be}{\beta}

\newcommand{\de}{\delta}
\newcommand{\om}{\omega}

\newcommand{\la}{\lambda}
\newcommand{\si}{\sigma}

\newcommand{\blist}{\begin{itemize}}

\newcommand{\elist}{\end{itemize}}

\providecommand{\href}[2]{#2}

\DeclareFontFamily{OT1}{rsfs}{}
\DeclareFontShape{OT1}{rsfs}{m}{n}{ <-7> rsfs5 <7-10> rsfs7 <10->rsfs10}{} 
\DeclareMathAlphabet{\mycal}{OT1}{rsfs}{m}{n}

\begin{document}
\begin{titlepage}

\renewcommand{\thefootnote}{\fnsymbol{footnote}}

\hfill TUW--02--15


\begin{center}
\vspace{0.5cm}

{\Large\bf Virtual Black Holes in Generalized Dilaton 
Theories}
\vspace{1.0cm}

{\bf D.\ Grumiller\footnotemark[1], W. Kummer\footnotemark[2], and D.V. Vassilevich\footnotemark[3]\footnotemark[4]
}
\vspace{7ex}

  {\footnotemark[1]\footnotemark[2]\footnotesize Institut f\"ur
    Theoretische Physik \\ Technische Universit\"at Wien \\ Wiedner
    Hauptstr.  8--10, A-1040 Wien, Austria}
  \vspace{2ex}

  {\footnotemark[3]\footnotesize Max-Planck-Institut f\"ur 
Mathematik in den Naturwissenschaften, 
\\
Inselstr. 22-26, D-04103 Leipzig,
Germany}

   \footnotetext[1]{e-mail: \texttt{grumil@hep.itp.tuwien.ac.at}}
   \footnotetext[2]{e-mail: \texttt{wkummer@tph.tuwien.ac.at}}
   \footnotetext[3]{e-mail: \texttt{vassil@itp.uni-leipzig.de}}
   \footnotetext[4]{On leave from V.Fock Institute of Physics,
St.Petersburg University, 198904 St.Petersburg, Russia}
\end{center}
\vspace{7ex}

\begin{abstract}

The virtual black hole phenomenon, which has been observed previously in 
specific models, is established for generic $2D$ dilaton gravity theories with
scalar matter. The ensuing effective line element can become asymptotically 
flat only for two classes of models, among them spherically reduced theories 
and the string inspired dilaton black hole (CGHS model).

We present simple expressions for the lowest order scalar field vertices of
the effective theory which one obtains after integrating out geometry exactly. 
Treating the boundary in natural and simple way asymptotic states, 
tree-level vertices and tree-level $S$-matrix are conformally invariant.

Examples are provided pinpointing the physical consequences of 
virtual black holes on the (CPT-invariant) $S$-matrix for gravitational 
scattering of scalar particles. For minimally coupled scalars the evaluation
of the $S$-matrix in closed form is straightforward. 

For a class of theories including the CGHS model all tree-graph vertices 
vanish, which explains the particular simplicity of that model and at the same 
time shows yet another essential difference to the Schwarzschild case.

\end{abstract}

PACS numbers: 
04.60.Kz; 04.60.Gw; 11.10.Lm; 11.80.Et

\vfill
\end{titlepage}

\section{Introduction}

Recent years have seen remarkable progress in the quantum treatment of 
two dimensional models of gravity, so-called Generalized Dilaton Theories 
(GDTs) (for a recent review cf. \cite{Grumiller:2002nm}), which include, most 
prominently, the Schwarzschild Black Hole (BH) and the string inspired dilaton 
BH \cite{Callan:1992rs}. 

GDTs have been quantized in the last decade mostly by Dirac's canonical method
\cite{Cangemi:1992bj}
and, less frequently, by the path integral technique
\cite{Haider:1994cw,Kummer:1998zs,Grumiller:2000ah,Fischer:2001vz,Grumiller:2001rg}. 
Indeed path integral quantization of a system with no propagating physical 
modes may appear to be something of an overkill, although both formalisms 
encounter essentially the same subtleties, albeit in different disguises.

However, once matter is switched on the path integral approach to us
appears to be superior. In addition to purely technical advantages the main 
reason is the much closer relation to genuine physical 
{\em observables}\footnote{
By {\em observable} we mean ``objects to be measured'' rather than observables 
in the formal sense of Dirac. Of course the latter play a crucial role in
Dirac quantization, but they are not necessarily identical to the former. 
Measuring the angular momentum of an electron (which is an observable 
in Dirac's sense) means to couple it to an external (magnetic) field and to
calculate a corresponding $S$-matrix element which is an {\em observable} in
the physical sense because it can be compared with experiments.} 
like $S$-matrix elements. In fact, results like the ones to be discussed in 
our present note have been derived until now exclusively within this 
theoretical setup\footnote{Of course, within the Dirac quantization approach
since the seminal work of Kucha\v{r} on the quantization of gravitational
waves \cite{Kuchar:1971xm} also interesting results with matter have been
obtained, e.g. in the thin shell approach 
\cite{Hajicek:2000mh}where bound states might occur 
\cite{Kouletsis:2001ma}.} \cite{Grumiller:2000ah,Fischer:2001vz}.

Especially as a consequence of the nonperturbative treatment of the geometric
part, achieved in a ``covariant Hamiltonian'' action, together with a specific
temporal gauge choice 
\cite{Haider:1994cw,Kummer:1998zs}, rather powerful new tools are 
available to tackle some persistent problems surrounding especially the 
Schwarzschild BH, but with consequences in other situations of General 
Relativity which can be reduced effectively to a two dimensional 
problem\footnote{At the quantum level the complications due to the 
``dimensional reduction anomaly'' 
\cite{Frolov:1999an} may come into play, but reasonable results for 
specific questions like the Hawking flux from BHs \cite{Kummer:1999zy} suggest 
that even these complications may be of limited importance.}.

Proceeding along well established paths of quantum field theory,
already at the level of the (nonlocal) vertices of matter fields, to 
be used in a systematic perturbative expansion in terms of Newton's constant, 
a highly nontrivial and physically intriguing phenomenon can be observed, 
namely the so-called  ``virtual black hole'' (VBH). This notion originally has 
been introduced by S. Hawking \cite{Hawking:1996ag}, but in our recent 
approach the VBH for spherically reduced gravity (SRG) emerges naturally in 
Minkowski signature space-time, without the necessity of additional {\em ad 
hoc} assumptions. 
The following observations are helpful to understand our notion of a VBH: the effective geometry that is obtained after taking into account the first nontrivial matter correction in our background independent quantization procedure contains a BH geometry localized on a light-like cut (see the Carter-Penrose diagram fig.\ \ref{fig:cp} below). The nontrivial part of this geometry looks like a BH geometry in a Rindler background. This type of BH is not felt by the asymptotic states but only indirectly via scattering processes (which are encoded in the lowest order vertices to be calculated in this work). It can be traced back to a discontinuity in the geometric part of a conserved quantity (denoted by ${\cal C}^{(g)}$) which exists in all theories under consideration, eq.\ (\ref{vbh:cg}) below. Thus, whenever ${\cal C}^{(g)}$ has a discontinuity a VBH is present.
The first discussion in \cite{Grumiller:2000ah} (SRG with in 
$D=2$ minimally coupled scalars) was too simple to yield an interesting 
$S$-matrix, unless mass terms of the scalar field were added. The situation 
had improved for non-minimally coupled scalars 
\cite{Fischer:2001vz} where the lowest order $S$-matrix 
indeed exhibited interesting features: forward scattering poles, monomial 
scaling with energy, CPT invariance, and pseudo-self-similarity in its 
kinematic sector \cite{Grumiller:2001rg}.

In the present work we extend this analysis to arbitrary GDTs and show that
the VBH phenomenon is generic. However, for one particular class of models, 
including the in the $2D$-community well-known CGHS model, no tree-level 
scattering exists. Therefore the VBH phenomenon is not observable there
(at least at the classical level).
This result again pinpoints its special role among GDTs and provides a 
somewhat physical way to explain its simplicity. It also confirms that the 
Schwarzschild BH and the dilaton BH differ essentially.

In order to make this note self-contained we recapitulate in section \ref{se:2}
some basic features of the path integral quantization. Section \ref{se:3} is 
devoted to the determination of the vertices which are already nonlocal at the 
tree level, generalizing previous results on SRG \cite{Grumiller:2000ah,
Fischer:2001vz} to generic dilaton theories (cf. 
(\ref{vbh:geometryaction}) or (\ref{vbh:sog}) below). Although the 
consequences of these vertices for tree level scattering of scalars
under the influence of their own gravitational interactions is a classical 
phenomenon, their treatment from an $S$-matrix point of view is advantageous. 
Even without an explicit evaluation of special cases the general features can 
be discussed quite broadly (section \ref{se:4}).

\section{Two dimensional quantum gravity with matter}\label{se:2}

In its first order gravity (FOG) version the GDT action reads
\eq{
L=L^{(FOG)}+L^{(m)}\,,
}{vbh:action} 
with the geometric part
\eq{
L^{(FOG)} = \int\limits_{\mathcal{M}_2} \, 
\left[ X_a D\wedge e^a + X d\wedge\omega + \epsilon \mathcal{V} 
(X^a X_a, X) \right]\,.
}{vbh:geometryaction}
Geometry is expressed by the one form zweibeine $e^a$ ($a=+,-$ in the local 
Lorentz frame using light-cone gauge) and the one form spin connection 
$\om^a{}_b=\eps^a{}_b\om$, appearing in the covariant derivative 
$D^a{}_b=\de^a_bd+\om^a{}_b$ (the volume form is denoted by 
$\epsilon=\frac{1}{2} \eps_{ab}e^a\wedge e^b$). The dependence on the 
auxiliary fields $X^a$ and the dilaton field $X$ in the potential
(using $X^aX_a=2X^+X^-$)
\begin{equation}
\mathcal{V}(X^a X_a, X) = X^+X^- U(X) +V(X) \label{vbh:calV}
\end{equation}
encodes the models relevant for our purposes.
The action (\ref{vbh:geometryaction}) is classically and quantum mechanically
equivalent \cite{Kummer:1998zs} to the more familiar general second order 
dilaton action
\begin{equation}
 L^{(SOG)}=\int d^2 x \sqrt{-g} \left[
-X\frac R2 -\frac{U(X)}2 (\nabla X)^2 +V(X) \right]\,. 
\label{vbh:sog}
\end{equation}
For the matter part we choose a (non-minimally coupled) massless scalar field:
\eq{
L^{(m)} = \frac{1}{2} \int\limits_{\mathcal{M}_2} \, F(X) d\phi \wedge 
\ast d\phi\,,
}{vbh:matteraction}
with an -- in principle arbitrary -- coupling function $F(X)$. In practice
the cases $F=\rm const.$ (minimal coupling) and $F\propto X$ (SRG) are the most
relevant ones.

The path integration can be performed using e.g. a ``temporal gauge''
\begin{equation}
\om_0=0,\quad e_0^-=1,\quad e_0^+=0\,.
\label{vbh:tempgauge}
\end{equation}
It is convenient to introduce comprehensive 
notations (in accordance with \cite{Grumiller:2002nm}) for
the remaining geometric variables:
\begin{equation} 
(q_1,q_2,q_3):=(\om_1,e^-_1,e^+_1)\,,\qquad
(p_1,p_2,p_3):=(X,X^+,X^-)\,.\label{vbh:notations}
\end{equation}
In terms of (\ref{vbh:notations}) the action (\ref{vbh:action}) can be 
rewritten as
\begin{eqnarray}
&&L=\int d^2 x\Big[ p_i\dot q_i +q_1p_2 -q_3\left(
V(p_1) +U(p_1)p_2p_3 \right) \nonumber \\
&&\qquad\qquad\quad +F(p_1) \left( (\partial_1\phi )(\partial_0\phi)
-q_2(\partial_0\phi )^2 \right) \Big] \,,\label{vbh:gfaction}
\end{eqnarray}
where the gauge condition (\ref{vbh:tempgauge}) has been taken into account.
The abbreviations
\begin{equation}
\msc_0:=\frac{1}{2}(\partial_0\phi)^2\,,\qquad 
\msc_1:=\frac{1}{2}(\partial_0\phi)(\partial_1\phi)\,.
\label{vbh:f0f1}
\end{equation}
are suggested by (\ref{vbh:gfaction}). 

The action (\ref{vbh:gfaction}) 
is linear in $q_i$. This is also true for the matter action 
(\ref{vbh:matteraction}) when it is expressed in terms of the variables 
(\ref{vbh:notations}) and for the source terms for $q_i$ which are neither 
included explicitly in (\ref{vbh:gfaction}) nor in what follows. Therefore,
the integration over $q_i$ can be performed exactly -- without the necessity
of introducing a split of geometry into ``background'' and ``fluctuations'' --
yielding three functional $\delta$-fuctions $\delta (p_i -p_i^{cl})$,
where $p_i^{cl}$ are solutions of the classical equations of motion for
$p_i$ in the presence of external sources.
For each specific model the $p_i^{cl}$ are given functions of $\msc_0$ 
and $\msc_1$. The three delta functions are then used to integrate $p_i$.  
Retaining only the dependence on the source $\sigma$ of the scalar 
field the simplified generating functional of Green functions reads\footnote{
In the present context questions regarding the measure, back reactions and 
geometric source terms are irrelevant. Therefore, 
the generating functional of 
Green functions simplifies considerably as compared to the exact case 
\cite{Kummer:1998zs,Fischer:2001vz,Grumiller:2002nm}.}
\begin{eqnarray}
&&W(\si ) = \int (\mathcal{D}\phi) \exp{i L^{\rm eff}},
\label{eq:4.52}\\
&&L^{\rm eff}=\int d^2x\left[2F(p_1^{\rm cl})\msc_1 +
\tilde{\mathcal{L}}(p_i^{\rm cl}) + \si\phi \right],
\label{eq:4.53a}\\
&&\tilde{\mathcal{L}}(p_1^{\rm cl}) = - \tilde{g} w'(p_1^{\rm cl})\,,
\label{eq:4.53}
\end{eqnarray}
where
\begin{equation}
w(p_1):=\int^{p_1}I(y)V(y)dy\,,
\label{vbh:w}
\end{equation}
with
\begin{equation}
I(p_1):=\exp{\int^{p_1}U(y)dy}.
\label{vbh:IQ}
\end{equation}
Here $\tilde g$ plays the role of an effective coupling. It will be fixed
below establishing the unit of length. Note that 
$p_1^{\rm cl}$ depends non-locally and non-polynomially on $\msc_0$. In the
absence of external sources it is determined by the classical equations of
motion (e.o.m.) (\ref{2.1}),(\ref{2.2}) below.

Up to this point all path integrals have been exact. In the
next step the quadratic terms in $\phi$, resp. the linear expressions in 
$\msc_0$ and $\msc_1$, are isolated in (\ref{eq:4.53a}). Terms of higher order 
in  $\phi$ are interpreted as vertices in a standard perturbation theory. 

A peculiar property of $2D$ gravity theories is the presence of an 
``absolute'' (in space and time) conservation law $d\mathcal{C}=d
(\mathcal{C}^{(g)}+\mathcal{C}^{(m)})=0$ 
\cite{Kummer:1992rt} 
even when matter is taken into account 
\cite{Kummer:1995qv}. 
Its geometric part only depends on the auxiliar fields, resp. $p_i$ (cf. 
(\ref{vbh:notations})). The (classical) expression for the latter
\begin{equation}
\mathcal{C}^{(g)}:=I(p_1)p_2p_3+w(p_1)\,,\label{vbh:Cg}
\end{equation}
will be needed below with $I$ of (\ref{vbh:IQ}) playing the role of an
integrating factor. $\mathcal{C}^{(g)}$ is related to the so-called 
``mass-aspect function'' \cite{Grumiller:1999rz} and becomes simply 
proportional to the ADM mass for asymptotically flat metrics. 

It should be noted that in models (\ref{vbh:sog}) which are connected by
a conformal transformation 
\eq{
g_{\mu\nu} = e^{2\rho (X)}\tilde{g}_{\mu\nu}\,,\hspace{0.5cm}
U(X)=\tilde{U}(X)+2\rho'(X)\,,\hspace{0.5cm}V=e^{-2\rho(X)}\tilde{V}(X)\,,
}{vbh:ct}
the combination $w(p_1)=\tilde{w}(p_1)$ as defined in (\ref{vbh:w}) is 
invariant.

\section{Construction of the lowest order tree-graphs}\label{se:3}

In the present paper we are interested  in the 
tree level amplitudes and in the generic properties of ensuing
$S$-matrix elements. The final result can be formulated in
a rather intuititive way as explained below. A more rigorous derivation
can be found in our previous work \cite{Grumiller:2000ah,Fischer:2001vz,Grumiller:2002nm}.

\begin{figure}
\epsfig{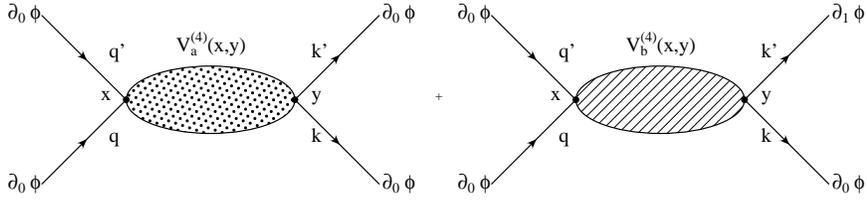}
\caption{The total $V^{(4)}$-vertex (with outer legs) contains a symmetric
contribution $V^{(4)}_a$ and (for non-minimal coupling) a non-symmetric one
$V^{(4)}_b$. The shaded blobs depict the intermediate interactions with VBHs.}
\label{fig1}
\end{figure}

The lowest order vertex from (\ref{eq:4.53}) is quartic in $\phi$. Thus in 
the notations (\ref{vbh:f0f1}) the generic four-point interaction term reads
\begin{multline}
V^{(4)}=\int d^2xd^2y {\Big[} V^{(4)}_a(x,y)\msc_0(x)\msc_0(y) +
V^{(4)}_b(x,y)\msc_0(x)\msc_1(y) \\
+ V^{(4)}_c(x,y)\msc_1(x)\msc_1(y) {\Big]}
\label{vbh:gvert}
\end{multline}
with some kernels $V^{(4)}_{a,b,c}(x,y)$. As $p_1^{cl}$ is independent of 
$\msc_1$ (cf. (\ref{2.1}), (\ref{2.2}) below) it is clear that the interaction 
term with $V^{(4)}_c$ never appears in $V^{(4)}$ which is graphically 
represented in fig. \ref{fig1}.

In principle $V^{(4)}$ could be abstracted directly from (\ref{eq:4.53a}). 
However, the expressions $p_i^{\rm cl}=p_i^{\rm cl}(\msc_0)$ require the
solution of three coupled first order differential equations which are nothing
else than the classical e.o.m. (\ref{2.1})-(\ref{2.3}) 
below. The geometric quantities $q_i$ could be obtained as expectation values. 
Also for them it is much preferable to go back to their (classical) e.o.m.\,. 
Then, for example, to calculate $V^{(4)}_{a}(x,y)$, one may use the following
recipe \cite{Kummer:1998zs,Grumiller:2000ah,Fischer:2001vz,Grumiller:2002nm}:
With the matter distribution localized at a point $y$
\begin{equation}
\msc_i(x)=c_i \delta^2 (x-y)\,,\hspace{0.5cm}i=0,1\,. 
\label{vbh:c0}
\end{equation}
one computes solutions $q_i(x;c_i,y)$, $p_i(x;c_i,y)$ of the classical e.o.m.
with the matter sources given by (\ref{vbh:c0}). Recalling that the 
coefficient in front of $\msc_0$ in (\ref{vbh:gfaction}) is $[-2F(p_1)q_2]$,
$V^{(4)}_{a}(x,y)$ simply is given by $[-2F(p_i(x;c_i,y))q_2(x;c_i,y)]$,
taken at linear order of $c_0$. The vertex $V^{(4)}_{b}(x,y)$ can be 
constructed in a similar manner. In fact, for all tree-graph vertices (with 
$2N$ outer legs) the knowledge of the solutions of the classical e.o.m. with 
localized matter sources (at $N-1$ different points) is sufficient.

\subsection{Equations of motion}

The gauge-fixed action (\ref{vbh:gfaction}) yields the following e.o.m.:
\begin{align}
\partial_0 p_1 &= p_2\,, \label{2.1} \\
\partial_0 p_2 &= -2F(p_1)\msc_0\,, \label{2.2} \\
\partial_0 p_3 &= -V(p_1)-p_2p_3U(p_1)\,, \label{2.3} \\
\partial_0 q_1 &= q_3(V'(p_1)+p_2p_3U'(p_1))-2F'(p_1)(\msc_1-q_2\msc_0)\,, 
\label{2.4} \\
\partial_0 q_2 &= -q_1+q_3p_3U(p_1)\,, \label{2.5} \\
\partial_0 q_3 &= q_3p_2U(p_1)\, \label{2.6}. 
\end{align}
The system (\ref{2.1})-(\ref{2.6}) can be solved exactly for $\msc_0=\msc_1=0$,
as well as for localized matter (\ref{vbh:c0}). Then two patches ($x^0<y^0$ 
and $x^0>y^0$) must be matched by appropriate conditions at $x^0=y^0$.
The structure of (\ref{2.1})-(\ref{2.6}) shows that
$p_1,p_3,q_2,q_3$ are continuous, whereas $p_2,q_1$ may jump at $x^0=y^0$. 
This yields six conditions relating the ``asymptotic'' integration constants 
($x^0>y^0$) to the ``inner'' ones ($x^0<y^0$). 

The conditions for the determination of the six asymptotic integration 
constants are:
\blist
\item $\left.p_1\right|_{x^0\to\infty} = x^0$ fixes two integration constants. 
It implies $\left.p_2\right|_{x^0\to\infty} = 1$.
\item $\left.{\mathcal C}^{(g)}\right|_{x^0\to\infty} = \mathcal{C}_\infty$ 
by (\ref{vbh:Cg}) yields the constant in $p_3$. We will choose 
$\mathcal{C}_\infty=0$, which corresponds to the ground state -- e.g. for SRG
which is asymptotically flat this condition yields Minkowski space-time.
\item $\left.q_3\right|_{x^0\to\infty} = 1\cdot I(p_1)$ establishes the unit 
of length.
\item The remaining two integration constants are called $m_\infty$ and 
$a_\infty$, because for SRG they correspond to a Schwarzschild and a Rindler
term, respectively. Both of them enter $\left.q_2\right|_{x^0\to\infty}$.
\elist
It should be noted that not all six integration constants are independent from 
each other: Eqs. (\ref{2.1})-(\ref{2.6}) are not the complete set of
(classical) e.o.m. of the model. Without gauge fixing also equations
from variation of $e_0^\pm$ and $\om_0$ -- the secondary
constraints in the Halitonian formalism \cite{Grumiller:2002nm} -- must hold. 
In the quantum formalism they appear as ``Ward identities'' and resemble
(\ref{2.1})-(\ref{2.2}), but with the $\partial_1$ derivative replacing 
$\partial_0$. It turns out that they imply two additional independent 
relations,
\eq{
\mathcal{C}_\infty = m_\infty\,, \hspace{0.5cm} a_\infty = 0\,.
}{vbh:icrel}
leaving the only non-trivial asymptotic quantity to be $m_\infty$. Due to our conventions $m_\infty$ is negative for a positive BH mass.

Fixing these integration constants in this manner the global solutions for the 
momenta from (\ref{2.1})-(\ref{2.3}) to ${\mathcal O}(c_0)$ are
\begin{align}
p_1 &= x^0+2F(y^0)(x^0-y^0)h_0\,, \label{vbh:p1} \\
p_2 &= 1+2F(y^0)h_0\,, \\
p_3 &= I^{-1}(p_1)\left(m_\infty-2F(y^0)w(y^0)h_0-\frac{1}{p_2}w(p_1)\right)\,,
\end{align}
with the abbreviations
\begin{equation}
h_i = c_i\theta(y^0-x^0)\delta(x^1-y^1)\,,\hspace{0.5cm} i=0,1\,.
\end{equation}
From now on all quantities will be given only to the required linear order in 
$c_0$ and $c_1$.
The geometric part of the conserved quantity from (\ref{vbh:Cg}) becomes
\eq{
{\mathcal C}^{(g)} = m_\infty+2F(y^0)\left(m_\infty-w(y^0)\right)h_0\,.
}{vbh:cg}
Thus, $m_\infty$ fixes the asymptotic value of ${\mathcal C}^{(g)}$. One can 
observe already the VBH-phenomenon: The discontinuity in $h_0$ at $x^0=y^0$ 
also carries over to $\mathcal{C}^{(g)}$. Thus the VBH is a generic feature of 
{\em all} GDTs. 

The solution for the component $q_1$ of the spin-connection is not needed for 
the line element and only $\partial_0 q_1$ (which can be read off from 
(\ref{2.4})) is needed for the curvature scalar\footnote{For SRG the curvature 
scalar can be calculated more elegantly using the Kerr-Schild decomposition
\cite{Grumiller:2001rg}.}.
The line element in the gauge in (\ref{vbh:tempgauge}) from 
(\ref{vbh:notations}) only depends on the zweibeine $q_2,q_3$ which are the 
solutions of (\ref{2.4}),(\ref{2.5}): 
\begin{align}
q_2 &= m_\infty-w(p_1)-2(x^0-y^0)F'(y^0)h_1  \nonumber \\
& \quad +\left[4F(y^0)\left(w(x^0)-w(y^0)\right)-2(x^0-y^0)\left.(Fw)'
\right|_{y^0}\right]h_0\,, \label{eq:q2} \\
q_3 &= I(p_1)\,. \label{eq:q3}
\end{align}

\subsection{Effective geometry}

The line element $(ds)^2=2e^+\otimes e^-$ in the gauge (\ref{vbh:tempgauge}) 
reads
\eq{
(ds)^2=2q_3dx^0dx^1+2q_2q_3(dx^1)^2\,.
}{vbh:ds0}
In order to bring it into a more familiar form we define a ``radial'' variable
\eq{
dr=b q_3(x^0)dx^0, \hspace{0.5cm}b\in\mathbb{R}^\ast\,,
}{vbh:radius}
and a ``null coordinate''
\eq{
du=b^{-1}dx^1\,,
}{vbh:null}
thus obtaining the line element in Sachs-Bondi form
\eq{
(ds)^2=2drdu+K(r,u;r_0,u_0)(du)^2\,,
}{vbh:ds1}
with the ``Killing norm''\footnote{The quantities $(r_0,u_0)$ are related to 
$(y^0,y^1)$ like $(r,u)$ to $(x^0,x^1)$.} $K(r,u;r_0,u_0)=2b^2q_2q_3$.

More explicitly $K$ reads
\meq{
K(r,u;r_0,u_0)=K_\infty \left[1+2F(y^0)U(x^0)(x^0-y^0)h_0\right] - 4b^2
I(x^0)(x^0-y^0)\\
\left[F(y^0)\left(V(x^0)I(x^0)+V(y^0)I(y^0)\right)h_0+F'(y^0)\left(w(y^0)h_0 + 
h_1\right)\right] \\
+8b^2I(x^0)F(y^0)\left(w(x^0)-w(y^0)\right) h_0\,, 
}{vbh:killing}
with
\eq{
K_\infty=2b^2I(x^0)\left.\left(m_\infty-w(x^0)\right)\right|_{c=0}\,.
}{vbh:kinfty}
Its continuity at $x^0=y^0$ is manifest in (\ref{vbh:killing}). For SRG in the 
VBH region a Schwarzschild term (proportional to $1/r$) and a Rindler term 
(proportional to $r$) are present in (\ref{vbh:killing}). Their appearance is 
somewhat surprising because apart from fixing the asymptotic boundary values 
of all geometric quantities we have made no assumption whatsoever on the 
geometry, except for its (Minkowskian) signature.

Depending on $I(x^0)$ and $w(x^0)$ asymptotic flatness\footnote{Statements
about asymptotic flatness are in general {\em not} invariant under conformal 
transformations because in addition to $w$-dependence we have dependence on
 the boundary value $m_\infty$ and also dependence on $I$, a conformally 
non-invariant quantity.\\ For definiteness we assume that the asymptotic 
region corresponds to $x^0,r\to\infty$.} can be achieved for which there are 
two possibilities:
\blist
\item $\left.I(x^0)w(x^0)\right|_{c=0} = {\rm const.} \in\mathbb{R}^\ast$ and 
$\lim_{r\to\infty} I(r) \propto r^\al$, with $\al \leq 0$.  
\item $I(x^0) = {\rm const.} \in\mathbb{R}^\ast$ and 
$\lim_{r\to\infty} w(r) \propto r^\al$, with $\al \leq 0$. 
\elist
The first scenario is fulfilled for all SRG models arising by reduction from
$D$ dimensions ($F \propto X$, $x^0 \propto r^{D-2}$, $D>3$, 
$\la\in\mathbb{R}^\ast$)
\begin{align}
U(X)&=-\frac{(D-3)}{(D-2)X}\,, \hspace{0.5cm}V(X)=-\frac{\la^2}{2}(D-2)(D-3)
X^{(D-4)/(D-2)}\,, \\ 
I(X)&=X^{(3-D)/(D-2)}\,, \hspace{0.5cm}w(X)=-\frac{\la^2}{2} (D-2)^2
X^{(D-3)/(D-2)}\,, 
\end{align}
including the CGHS model as the formal limiting case $D\to\infty$, $F = \rm 
const.$ and $\la \propto 1/D$ \cite{Grumiller:2002nm}. The second case applies 
e.g. when SRG models are transformed conformally so that $\tilde{U}=0$.

For GDTs the only independent physical geometric quantity is 
the curvature scalar\footnote{There is a slight subtlety at this point, as there are two possible definitions of the curvature scalar (one with torsion, which is the one we are using here and one without torsion): in conformal frames where $I\neq 1$ the curvature scalar (as defined by the Hodge dual of the curvature two form) is not just the second derivative of the Killing norm, because part of the geometric information is encoded in the torsion. In the absence of matter one obtains for the curvature as derived from the Killing norm
\eq{
R=2 \left(V'(p_1)+p_2p_3U'(p_1)\right)-4\frac{w''(p_1)}{I(p_1)}\,,
}{err1}
while curvature as defined as the Hodge dual of the curvature 2-form reads
\eq{
R=-2\left(V'(p_1)+p_2p_3U'(p_1)\right)\,.
}{err2}
In the absence of torsion ($U=0$) both definitions coincide. The torsion 2-form $T^a$ is proportional to the volume 2-form times $U(p_1)t^a$ with $t^+=I(p_1)$ and $t^-={\cal C}^{(g)}-w(p_1)$.} $R^{\rm (VBH)} = 2 (e)^{-1} \tilde{\epsilon}^{\al\be}
\partial_\al\om_\be = -2 q_3^{-1} \partial_0 q_1$
which in terms of the solutions  (\ref{eq:q3}) and (\ref{2.4}) becomes
\eq{
R^{\rm (VBH)} = -2 \left(V'(p_1)+p_2p_3U'(p_1)\right) 
+ 4(I(p_1))^{-1}F'(p_1) \left(\msc_1-q_2\msc_0\right)\,.
}{vbh:ricci2} 
One can read off from  (\ref{vbh:ricci2}) that the first term provides a step 
function at $x^0=y^0$ while the second term yields a contribution\footnote{In 
some models (e.g. all SRG models) an additional ($\de$-like) contribution at 
the ``origin'' $x^0=0$ may arise \cite{Balasin:1994kf}.} $\de(y^0-x^0)$.
The latter is absent for minimal coupling. If additionally $U'=0$ -- as for
instance in the Katanaev-Volovich model 
\cite{Katanaev:1986wk} -- the curvature scalar becomes even continuous. As the 
curvature depends on $U$ and $V$ separately it is quite different in 
conformally related theories.
\begin{wrapfigure}{r}{3cm}
\epsfig{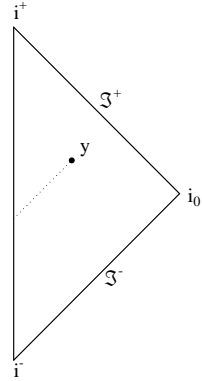}
\caption{CP diagram of the VBH}
\label{fig:cp}
\end{wrapfigure}
The Carter-Penrose (CP) diagram of the VBH geometry is analogous to the SRG 
case discussed in ref. \cite{Grumiller:2001rg}: there will be a light-like cut
where the non-trivial part of geometry (namely the VBH) is localized with
$\de$-like contributions to the curvature scalar on its end-points. The rest
of the CP-diagram is determined by (\ref{vbh:kinfty}). For instance, if the
potentials and the asymptotic mass $m_\infty$ are chosen such that $K_\infty$
is constant, then the CP diagram is given by the one for Minkowski space beside
the VBH. One can simply determine the full CP-diagram by disregarding the VBH
in a first step (thus drawing an ordinary CP-diagram for a line element
(\ref{vbh:ds1}) with $K$ being replaced by $K_\infty$) and then simply adding
the light-like cut representing the VBH. 

As in the previous case one obtains a family of CP-diagrams labelled by the
endpoint (denoted by $y$ in fig. \ref{fig:cp}). All these (off-shell) 
geometries are (continuously) summed in the $S$-matrix.


\subsection{Symmetric four-point vertex}

To obtain the interaction vertex $V_a^{(4)}$ one has\footnote{Especially 
at this point it might be convenient to consult 
section 8 of the review \cite{Grumiller:2002nm}
or corresponding parts of the original papers \cite{Kummer:1998zs,
Grumiller:2000ah,Fischer:2001vz}.} to  substitute the classical solutions 
(\ref{vbh:p1})-(\ref{eq:q3}) in the interaction term $-F(p_1)q_2 \msc_0(x)$,
truncate it to linear order in $c_0$, multiply by $\msc_0(y)$, and integrate
over the space-time.
\meq{
V_{a}^{(4)}=-2\int_x\int_y \msc_0(x)\msc_0(y)\theta(x^0-y^0)\de(x^1-y^1)F(x^0)
F(y^0)\\
\Bigg[4\left(w(x^0)-w(y^0)\right)-2(x^0-y^0)\Big(w'(x^0)+w'(y^0)\\
+\frac{F'(y^0)}{F(y^0)}w(y^0)+\frac{F'(x^0)}{F(x^0)}\left(w(x^0)-m_\infty
\right)\Big)\Bigg]\,.
}{vbh:sym}
This vertex has the following properties:
\blist
\item It vanishes for $x^0=y^0$.
\item It depends only on the combination  (\ref{vbh:w}) of $U$ and $V$
in the function $w(p_1)$; since $w(p_1)$ is invariant under 
conformal transformations also the vertex is invariant, if $m_\infty$ was 
fixed to the same  value in all conformal frames. 
\item It respects the $\mathbb{Z}_2$ symmetry $F(p_1)\to-F(p_1)$; thus, a 
change of the sign of the gravitational coupling constant (implicitly 
contained in $F(p_1)$) does not change the result for this vertex\footnote{
It is interesting to note that the one-loop quantum effective action
for non-minimally coupled matter fields respects a different 
$\mathbb{Z}_2$ symmetry \cite{Vassilevich:2000kt}:
$F(p_1)\to F(p_1)^{-1}$.}.
\elist

\subsection{Non-symmetric four-point vertex}

An additional vertex arises for non-minimal coupling ($F'\neq 0$ in 
(\ref{vbh:matteraction})): 
\eq{
V_{b}^{(4)}=-4\int_x\int_y \msc_0(x)\msc_1(y)\de(x^1-y^1)F(x^0)F'(y^0)
\left|x^0-y^0\right|\,.
}{vbh:asy}
It shares the properties with its symmetric counterpart, except for
the fact that it is even independent of $U$ and $V$.

No further vertices appear, unless one adds mass-terms (they yield additional 
vertices as shown in ref. \cite{Grumiller:2000ah}) or local self-interactions 
(which are rather trivial).

\section[$S$-matrix elements]{$\boldsymbol{S}$-matrix elements}\label{se:4}

The $S$-matrix element with ingoing modes $q$, $q'$ and outgoing ones 
$k$, $k'$,
\eq{
T(q, q'; k, k') = \frac{1}{2} \left< 0 \left| a^-_ka^-_{k'} \left(V^{(4)}_a 
+ V^{(4)}_b \right) a^+_qa^+_{q'}\right| 0 \right>\,, 
}{vbh:smatrix}
depends not exclusively on the vertices (\ref{vbh:sym}), (\ref{vbh:asy}), but 
also on the asymptotic states of the scalar field with corresponding 
creation/anihilation operators $a^\pm$ obeying canonical commutation 
relations of the form $\left[a^-_k,a^+_{k'}\right]\propto \de(k-k')$.

Their explicit form is model dependent and sensitive to $m_\infty$. That is
why from now on we restrict ourselves to general statements rather than
calculations which have been already rather lengthy in previous work 
on SRG \cite{Fischer:2001vz}.

\subsection{Asymptotic matter states}

The metric extracted from $(ds)^2$ in (\ref{vbh:ds0}) at $x^0\to\infty$
determines the asymptotic matter states in the $S$-matrix. They not only 
depend on the model but also on the value of $m_\infty$. In this limit the 
Klein-Gordon equation, expressed by the asymptotic form of 
$q_2=m_\infty-w(x^0)$ reads
\eq{
\partial_1\partial_0 \phi + \frac{F'(x^0)}{2F(x^0)}\partial_1 \phi =
q_2\partial_0\partial_0 \phi + q_2 \frac{F'(x^0)}{F(x^0)}\partial_0 \phi 
-w'(x^0)\partial_0 \phi\,,
}{vbh:matter}
If (\ref{vbh:matter}) can be solved exactly and the set of solutions is
complete and normalizable (in an appropriate sense) a Fock space for incoming 
and outgoing scalars can be constructed in the usual way. 
Eq. (\ref{vbh:matter}) is conformally invariant with the previous {\em caveat} 
concerning $m_\infty$. Since the integration constant $m_\infty$ 
enters (\ref{vbh:matter}) the asymptotic states depend on the choice 
of boundary conditions. 

Most string inspired dilaton models exclusively use minimally coupled scalars
($F(p_1)={\rm const.}$). In that case (\ref{vbh:matter}) simplifies drastically:
\eq{
\partial_0(\partial_1\phi - q_2 \partial_0 \phi)=0
}{vbh:1}
Solutions of this equation are outgoing modes
\eq{
\phi_{\rm out} = a_k^+ \exp{\left(ikx^1\right)}\,,
}{vbh:2}
and ingoing ones
\eq{
\phi_{\rm in} = a_k^- \exp{\left(-ik\left(x^1+\int^{x^0}\frac{dz}{m_\infty
-w(z)}\right)\right)}\,.
}{vbh:3}
Using first (\ref{vbh:radius}) and (\ref{vbh:null}) and then the 
Regge-Wheeler redefinition $dt=du+dr/K_\infty(r;r_0)$ with $K_\infty$ 
given by (\ref{vbh:kinfty}) one can switch to the more transparent $(r,t)$ 
coordinates. In terms of these we obtain ordinary asymptotic plane waves.
After defining a properly normalized commutation relation between
$a_k^-$ and $a_k^+$ one can build the Fock space and calculate the $S$-matrix 
using the vertex (\ref{vbh:sym}) in complete analogy to 
\cite{Grumiller:2000ah} where the special case $m_\infty=0$,
$U(p_1)=-1/(2p_1)$ and $V(p_1)={\rm const.}$ has been investigated.

We emphasize that only one vertex is present in the minimally coupled case.
For the model considered in \cite{Grumiller:2000ah} this has led to the 
conclusion that one of the following three alternatives must hold:
1. the scattering amplitude diverges, 2. the scattering amplitude vanishes (if 
the virtual black hole is plugged by an {\em ad hoc} regularity condition) or
3. the scalar field acquires a mass term, thus producing a second four-point 
vertex.

Considering non-minimally coupled scalars induces two important complications: 
firstly, the asymptotic Klein-Gordon equation (and hence also the asymptotic 
states) differs and secondly, an additional vertex (\ref{vbh:asy}) exists. For
SRG especially the second drawback turned into a virtue, since the scattering
amplitude exhibited some very nice features (to start with, it was finite
without additional regularity conditions as opposed to the minimal case)
\cite{Fischer:2001vz,Grumiller:2001pt,Grumiller:2001rg}.

\subsection{CPT invariance}

Since our effective theory is non-local, CPT-invariance is not guaranteed by
the CPT theorem\footnote{The parent dilaton gravity theory is local but
contains singular interactions. 
} \cite{Streater:1989vi}. Therefore, possibilities of CPT 
violation must be explored because they could imply a preferred
direction of time.

Indeed, the result obtained for SRG in 
\cite{Grumiller:2001rg} can be generalized straightforwardly to arbitrary 
dilaton models. There is, however, a subtlety: in those previous calculations
$m_\infty=0$ was a natural consequence of the asymptotically flat metric. Under
this assumption, the gauge choice $\om_0=0$, $e_0^-=-1$, $e_0^+=0$ led to
some trivial sign changes in intermediate formulae (in particular, $e_1^-$ 
flipped its sign) and an overall sign change in the (real) 
scattering amplitude. From this and the fact that $C$ and $P$ act trivially
$CPT$ invariance could be established. 

In the presence of a non-vanishing integration constant $m_\infty$ 
the question arises how to relate its values for different gauges.
There is only one way to retain $CPT$ invariance: the new value for
$m_\infty$ must be related to the old one by  $m_\infty^{CPT}=-m_\infty$. Then
$e_1^-$ (and also the vertices and the amplitude) again simply flip their sign.

The physical interpretation of this treatment of the integration constant is
model dependent. For asymptotically flat models it means that the space-time 
and its mirror version must have the same ADM mass. For instance, in SRG
the term $1-2m_\infty/r$ present in the Killing norm has to change into 
$-1+2(-m_\infty^{CPT})/r$ and thus $m_\infty^{CPT}=-m_\infty$; note that the 
ADM mass is $+m_\infty$ in the first case and $-m_\infty^{CPT}$ in the second
case, so (by construction) it does not change. For generic other models no 
such interpretation seems to be available.

\subsection{Conformal invariance}

\subsubsection{General considerations}

As can be seen directly from (\ref{vbh:ricci2}) the curvature scalar 
obviously is not invariant under conformal transformations. For GDTs without 
matter (but with coupling to test particles because otherwise geometry is 
void\footnote{Geometry without test-particles to probe it is like the fiction 
of ``empty space'' or the fiction of quantum mechanical ``observables''
without external sources to probe them \cite{Grumiller:2000hp}.}) this leads 
to the immediate conclusion that GDTs are {\em not} 
conformally invariant -- for instance, the global structure can change by 
conformal transformations because in general they have one or more singular 
points (cf. e.g. \cite{Grumiller:2000hp} and references therein).

In the present context, however, we do not need test-particles anymore, because
we have (scalar) matter available to test geometry, e.g. by preparing some
initial scalar field configuration and measuring the final one\footnote{We
note parallels to the concept of ``quantum singularities'', where 
test-particles are replaced by a wavefunction and singularities can be
smoothed out 
\cite{Horowitz:1995gi}.}. The $S$-matrix 
describes the physical content of scattering processes.

Since both the asymptotic states and the vertices depend only on $w$ (a 
conformally invariant combination of the potentials $U$ and $V$), $F$ (the
coupling function, which by assumption is not changed by conformal 
transformations), and $m_\infty$ (the asymptotic mass-scale, which can be fixed
to the same numerical value in each conformal frame) we can conclude that
at tree level to lowest order in the scalar field conformal invariance exists. 
Classically one can trivially extend this result to the 
non-perturbative level (cf. eqs. (\ref{eq:4.53a}) and 
(\ref{eq:4.53}) which only depend on $w'(p_1)$), but {\em not} to 1-loop level 
because the conformal anomaly will destroy this invariance.

\subsubsection{The special role of CGHS}

%

 In the CGHS model ($U_{\rm CGHS}=-1/X,\, V_{\rm CGHS}=-2\la^{2}X,\, 
 F_{\rm CGHS}=\rm const.$)
 the derivative of the quantity (\ref{vbh:w}) is constant.
 Therefore, according to (\ref{vbh:sym}) $V^{\left( 4\right) }$ vanishes.
 Actually the absence of classical scalar vertices, to be used in perturbation
 theory, extends to arbitrary orders in $\msc_0$ and $\msc_1$
 of (\ref{vbh:f0f1}) as can be verified by inspection of the Lagrangian 
 (\ref{eq:4.53a}) with (\ref{eq:4.53}). Thus,
 nontrivial scattering with an arbitrary number of external scalar legs
 would have to result from higher order quantum effects, to be derived e.g.
 from quantum back reaction which has not been considered in the simplified
 path integral (\ref{eq:4.52}).
 The absence of classical scattering from the string inspired CGHS model
 explains its simplicity and, at the same time, questions its status as
 an emblematic BH laboratory relative to all other GDTs (including especially
 SRG itself) where $w'\left( p_{1}\right) \neq \rm const.$

 Obviously this ``scattering triviality'' extends to a wider class of GDTs 
 with minimally coupled scalars: if the potentials $V_{ST}$ and $U_{ST}$ in 
 (\ref{vbh:geometryaction}) or (\ref{vbh:sog}) are related by 
\eq{
U_{ST}(p_1)V_{ST}(p_1)+V'_{ST}(p_1)=0
}{vbh:st}
 the crucial quantity $w'_{ST}$ remains constant. 
As an example of such models the ``$ab$-family'' $U=-a/X$, $V\propto X^{(a+b)}$
\cite{Katanaev:1996bh} can be considered. The triviality condition 
(\ref{vbh:st}) establishes $b=0$, containing the class of soluble models 
studied by Fabbri and Russo \cite{Fabbri:1996bz}. As can be seen from fig. 
3.12 in \cite{Grumiller:2002nm} the value $b=0$ is a critical one for it 
separates different ``phases'' of Carter--Penrose diagrams.

\subsubsection{Presence of the VBH phenomenon}

 The VBH effect was tied to the appearance of a discontinuity in 
 $\mathcal {C}^{\left( g\right)}$ of (\ref{vbh:cg}).
 Again the conformally invariant quantity $w$, here combined with the constant 
 $m_{\infty }$, plays a central role. By the same token as above the VBH is
 conformally invariant. 

 Despite the presence of the VBH in the CGHS model we register the peculiar
 situation that it has no (classically) 
 observable consequences due to ``scattering triviality'' (\ref{vbh:st}).
Thus, scattering non-triviality implies the VBH phenomenon but not vice versa.
Only for ``generalized teleparallel\footnote{Teleparallel
theories obey $U=\rm const.$, $V=0$ while ``generalized'' refers to arbitrary
$U(p_1)$ but retaining $V=0$.}'' theories with $w(p_1)=w_0=\rm const.$ the
VBH phenomenon will be absent provided $m_\infty=w_0$.

\section{Conclusion and Outlook}\label{se:5}

We have established the existence of the virtual black hole phenomenon for all
Generalized Dilaton Theories with (scalar) matter, with the notable exception 
of ``generalized teleparallel'' theories. 
The corresponding effective geometry (\ref{vbh:ds1}) shows essentially the 
same features as the previously discussed case of spherically reduced gravity 
\cite{Grumiller:2000ah,Fischer:2001vz,Grumiller:2001rg}.
The ensuing tree-level $S$-matrix for scattering of scalars is CPT and 
conformally invariant. For the CGHS model and a class of similar ones
(\ref{vbh:st}) where $w'(p_1)$ as defined in (\ref{vbh:w}) is constant 
no tree-level vertices with any number of scalar legs are produced.

One can apply our methods to evaluate the $S$-matrix for all models where the 
set of solutions of (\ref{vbh:matter}) is complete and normalizable, i.e. 
where an asymptotic Fock space can be constructed. In some cases the results
will be not very illuminating \cite{Grumiller:2000ah}, in other cases they 
exhibit interesting properties \cite{Fischer:2001vz,Grumiller:2001rg}. For 
instance, spherically reduced gravity from arbitrary
dimension $D$ could be worthwhile studying, since on the one hand for $D=4$ we 
already know the result to be highly non-trivial and it would be rewarding to 
verify whether this is generic or rather a special feature of $D=4$. On the 
other hand the formal limit $D\to\infty$ will yield the CGHS model with 
{\em nonminimally} coupled scalars. Moreover, for $m_\infty=0$ the Fock space 
construction is straightforward in these particular cases, the asymptotic 
modes being $D$-dimensional $s$-waves.

Although some of our statements like tree-level conformal invariance and
``scattering triviality'' in certain models are valid for tree-vertices with 
any number of scalar legs, we were concerned mostly with the lowest order of 
these vertices ((\ref{vbh:sym}), (\ref{vbh:asy})). The non-symmetric
one vanishes for minimal coupling $F(p_1)=\rm const.$. 

The next obvious 
step at tree-level is a generalization to arbitrary 6-point vertices. They 
will be the first ones to break the $\mathbb{Z}_2$-symmetry discussed in the
paragraph below (\ref{vbh:sym}).
Loop calculations, which will give insight into the nature of the information
paradox are in progress. First results for the CGHS model are already available: the otherwise trivial specific heat becomes mass dependent at 1-loop level \cite{Grumiller:2003mc}.

We conclude this paper by noting that virtual black holes may be significant
already at the collider energies according to some models (cf. e.g.
\cite{Uehara:2002cj}). In principle, our analysis may be extended even
to creation of real black holes at future colliders 
\cite{Giddings:2001bu}. However, 
in a theoretical analysis one would have to give up spherical symmetry, so that a direct application of our approach would no longer be possible.

\section*{Acknowledgement}
 
This work has been supported by project P-14650-TPH of the Austrian Science 
Foundation (FWF) and the MPI MIS (Leipzig). We thank P. Fischer
for discussion. D.G. is grateful to D. Konkowski for drawing his attention to 
the concept of quantum singularities.


\end{document}